\RequirePackage{lineno}
\documentclass[prd,twocolumn,showpacs,preprintnumbers,floatfix,amsmath,amssymb,amsfonts]{revtex4}

\usepackage{graphicx}
\usepackage{amsmath}
\usepackage{dcolumn}
\usepackage{epsfig}
\usepackage{psfrag}
\usepackage{subfigure}
\usepackage{hyperref}

\RequirePackage{xspace}

\def\TeV{\ifmmode {\mathrm{\ Te\kern -0.1em V}}\else
	                   \textrm{Te\kern -0.1em V}\fi}%
\def\GeV{\ifmmode {\mathrm{\ Ge\kern -0.1em V}}\else
	                   \textrm{Ge\kern -0.1em V}\fi}%
\def\MeV{\ifmmode {\mathrm{\ Me\kern -0.1em V}}\else
	                   \textrm{Me\kern -0.1em V}\fi}%
\def\keV{\ifmmode {\mathrm{\ ke\kern -0.1em V}}\else
	                   \textrm{ke\kern -0.1em V}\fi}%
\def\eV{\ifmmode  {\mathrm{\ e\kern -0.1em V}}\else
	                   \textrm{e\kern -0.1em V}\fi}%
\let\tev=\TeV
\let\gev=\GeV

\def\TeVc{\ifmmode {\mathrm{\ Te\kern -0.1em V}/c}\else
	                   {\textrm{Te\kern -0.1em V}/$c$}\fi}%
\def\GeVc{\ifmmode {\mathrm{\ Ge\kern -0.1em V}/c}\else
	                   {\textrm{Ge\kern -0.1em V}/$c$}\fi}%
\def\MeVc{\ifmmode {\mathrm{\ Me\kern -0.1em V}/c}\else
	                   {\textrm{Me\kern -0.1em V}/$c$}\fi}%
\def\keVc{\ifmmode {\mathrm{\ ke\kern -0.1em V}/c}\else
	                   {\textrm{ke\kern -0.1em V}/$c$}\fi}%
\def\eVc{\ifmmode  {\mathrm{\ e\kern -0.1em V}/c}\else
	                   {\textrm{e\kern -0.1em V}/$c$}\fi}%
	
\def\cm{\ifmmode  {\mathrm{\ cm}}\else	                   
\textrm{~cm}\fi}%

\def\mm{\ifmmode  {\mathrm{\ mm}}\else	                   
\textrm{~mm}\fi}%
	
\usepackage{relsize}
\def\babar{\mbox{\slshape B\kern-0.1em{\smaller A}\kern-0.1em
    B\kern-0.1em{\smaller A\kern-0.2em R}}}

\begin{document}

\title{Identifying the $b$ quark inside a boosted hadronically decaying top quark \\using jet substructure in its  center-of-mass frame}

\author{Chunhui Chen} 
\affiliation{Department of Physics and Astronomy, Iowa State University, Ames, Iowa 50011, USA}

\begin{abstract}
In this paper we study the identification of the $b$ quark inside a boosted hadronically decaying top quark
in the center-of-mass frame of the jet.
We demonstrate that the method can be used to greatly reduce the QCD jet background even in a very high pileup condition.
The method has a much smaller fake rate  for QCD jets compared to typical $b$ quark
identification algorithms in jets at the same signal efficiency.  
When combining the $b$ quark identification in the center-of-mass frame of the jet with  jet substructure information,
 we can improve the rejection rate of QCD jet background by almost an order of magnitude 
 while maintaining the same identification efficiency for the boosted top quark. 
\end{abstract}

\maketitle

Many new physics (NP) extensions beyond the standard model (SM) predict new heavy resonances with 
masses at the TeV scale.  Some of these heavy resonances, such  as a  new heavy gauge boson 
$Z^\prime$ or Kaluza-Klein  gluons from the bulk Randall-Sundrum model, or a right-handed charged gauge boson
$W^\prime_R$, can predominantly decay to a final state~\cite{Agashe:2006hk} containing top quarks.
Searches for new heavy resonances decaying to top quark final states  have been very activily 
pursued to look for NP by the ATLAS and CMS experiments at the LHC~\cite{Aad:2012wm,Aad:2012dpa,Aad:2012raa,Aad:2012ej,Chatrchyan:2012ku,Chatrchyan:2012cx,Chatrchyan:2012yca,Chatrchyan:2012gqa}.
Because the top quarks from the heavy resonance decay are highly boosted,  their hadronically decaying  products are so 
collimated that they are often reconstructed as single jets in the experiments.
In this paper, we define a hadronically decaying top as the top quark for which the $W$ boson daughter decays hadronically,
hereafter referred as a $t$ jet. Although the invariant mass of the reconstructed
jet ($m_{\rm jet}$) can be used to identify the $t$ jets from QCD jets, where the QCD jets are defined as those jets initiated
by a quark other than top or gluon, it does not provide enough discriminating power to
effectively distinguish $t$ jets from the overwhelming QCD background in many analyses.
Techniques based on jet substructure information~\cite{Thaler:2008ju,Kaplan:2008ie,Almeida:2008tp,Krohn:2009wm,Plehn:2010st,Chekanov:2010vc,Bhattacherjee:2010za,Rehermann:2010vq,Chekanov:2010gv,Jankowiak:2011qa,Thaler:2011gf,Soper:2012pb,Chen:2013ola} have been developed as additional experimental handles to  identify boosted hadronically 
decaying top quarks.

Since the top quark decays almost exclusively to a $W$ boson and $b$ quark final state, identifying the
$b$ quark from the top decay by exploring its long lifetime can provide additional distinguishing power
for the boosted hadronically decaying top quark. While the identification of isolated jets stemming from the hadronization of 
$b$ quarks ($b$-tagging) has been widely used in many experimental
measurements, its application in boosted hadronic top decay is more difficult because the charged tracks 
associated with the $b$ quark need to be disentangled from the ones generated by the $W$ boson.
In this paper, we extend the studies presented in Refs.~\cite{Chen:2011ah,Chen:2013ola} to explore the identification of
the $b$ quark inside a $t$ jet in the center-of-mass frame of the jet. 
We demonstrate that the method can greatly reduce the QCD jet background while maintaining a high
identification efficiency of the boosted top quark even in an environment with very large numbers
of multiple interactions per event (pileup). 

We use boosted $t$ jets, from the SM process of a top-antitop pair
($t\bar{t}$) production,  as a benchmark to study the identification of the $b$ quark inside. 
We only consider the background from the SM dijet production because its cross section 
is several orders of magnitude larger than those of other SM backgrounds.

All the events used in this analysis are produced using the P{\footnotesize ythia} 6.421 event generator~\cite{Sjostrand:2006za}
for $pp$ collisions at $14\,\rm TeV$ center-of-mass energy. The spread of the beam interaction point is assumed 
a Gaussian distribution with a width of $45\,(0.025)\,\mm$  in the longitudinal (transverse) direction~\cite{Aad:2008zzm}.
 In order to evaluate the performance of the proposed 
$b$ quark identification method with the currently expected experimental conditions at the LHC,
we generate Monte Carlo events with different average numbers of pileup~\cite{Sjostrand:2006za}
and then repeat our studies for each scenario.  
To simulate the finite resolution of the
calorimeter detector at the LHC, we divide the $(\eta, \phi)$ plane into $0.1\times 0.1$ cells. We sum over the energy
of particles entering each cell in each event, other than the neutrinos and muons, and assume a massless pseudoparticle,
also referred to as an energy cluster that has the same energy and points to the center of the  cell. These energy clusters are fed into the 
F{\footnotesize astJet} 3.0.1~\cite{Cacciari:2005hq} package for  jet reconstruction.
The jets are reconstructed using the anti-$k_T$ algorithm~\cite{Cacciari:2008gp}  
with a distance parameter of $R=0.6$. The  anti-$k_T$ jet algorithm is the default one used at the ATLAS and CMS experiments. 
As for the charged tracks, their momentum and vertex positions are smeared according to the expected resolutions of the ATLAS detector~\cite{Aad:2008zzm}.

We select jets with $p_{\rm T}\ge600\,\gev$ and $|\eta|\le1.9$ as $t$ jet candidates,
where $p_{\rm T}$ and $\eta$ are the transverse momentum and pseudorapidity of the jet, respectively.
We further require that the $t$ jet candidates have $50\,\gev\le m_{\rm jet}\le 350\,\gev$. 
All the $t$ jet candidates in an event are kept for further analysis.
For $b$-tagging, only charged tracks with $p_{\rm T} >1\,\gev$ and $|\eta|<2.5$ are considered. They
are also required to satisfy the criteria that  $|d_0|<1\,\mm$ and $|z_0-z_{\rm pv}|\sin\theta<1.5\,\mm$,
where $d_0$ and $z_0$ are the transverse and longitudinal impact parameter of the charged track, 
$z_{\rm pv}$ is the longitudinal position of the primary vertex, and $\theta$ is the polar angle of the
charged track. A  charged track is considered to be associated with a jet if the distance parameter of $\Delta R$ between the
track and the jet is less than $0.6$.

We define the center-of-mass frame (rest frame) of a jet as the frame where the four-momentum of the
jet is equal to $p^{\rm rest}_{\mu}\equiv (m_{\rm jet}, 0, 0, 0)$. The distribution of pseudoparticles of 
a boosted $t$ jet  in its center-of-mass frame
has a three-body decay topology as in the top quark rest frame. We recluster the energy clusters
of a jet to reconstruct subjets in the jet rest frame using a modified 
$e^+e^-$ Cambridge jet reconstruction algorithm~\cite{Dokshitzer:1997in}.
The algorithm performs sequential recombination of the pair of psedoparticles that is closest in angle $\Theta$,
except for $\Theta > 0.6$, where $\Theta$ is defined as the angle between two pseudoparticles in the jet rest frame.
The implementation of the modified $e^+e^-$ Cambridge jet algorithm is done by replacing the distance parameter of 
the existing $e^+e^-$ Cambridge jet algorithm in the F{\footnotesize astJet} 3.0.1~\cite{Cacciari:2005hq} package 
with the new choice of the distance parameter $\Theta$.
We only retain jets that have at least three subjets each with energy $E_{\rm jet}>10\,\gev$ in the $t$ jet reset frame. 
In the ideal situation with no pileup effects, this
requirement rejects approximately 60\,\% of the QCD jets, while keeping almost all the signal $t$ jets.
However, the rejection power drops significantly when the average number of multiple interactions per event increases
to 50 (100), in which more than 70\,\% (90\,\%) of the QCD jets
have at least three subjets with $E_{\rm jet}>10\,\gev$. 
Currently the maximum average number of pileup at LHC is slightly less than 20;
with expected higher energy and luminosity in the future, it is expected to reach 50, and even 
100 in the worst case scenario.

The most straightforward way to identify the $b$ quarks inside $t$ jets is to apply existing 
$b$-tagging algorithms in a jet directly. In this paper we study the tagging algorithms based on
charged track impact parameters as the algorithms are  widely used in many experiments. They are also
among the official $b$-tagging methods used by the ATLAS experiment~\cite{Aad:2009wy}. The impact parameters of tracks are computed
with respect to the primary vertex. They typically have significant nonzero values for the charged tracks
from the $b$ hadron decays because of its long lifetime. The impact parameter is signed to further discriminate the
tracks from $b$-hadron decay from tracks originating from the primary vertex based on the fact that the decay position 
of the $b$ hadron lies along its flight path. The sign of transverse impact parameter $d_0$ is determined using the
jet momentum $\vec{p}_{\rm jet}$, the track momentum $\vec{p}_{\rm trk}$ at the point of the closest approach $\vec{x}_{\rm trk}$~\cite{Aad:2009wy}
to the primary vertex position $\vec{x}_{\rm pv}$:
\begin{equation}
{\rm sign}(d_0)=(\vec{p}_{\rm jet}\times \vec{p}_{\rm trk})\cdot (\vec{p}_{\rm trk}\times (\vec{x}_{\rm pv}-\vec{x}_{\rm trk})).
\end{equation}
The sign of longitudinal impact parameter $z_0$ is measured by the sign of $(\eta_{\rm jet}-\eta_{\rm trk})\times z_{0,{\rm trk}}$,
where $\eta_{\rm jet}$ is the pseudorapidity of the jet, and $\eta_{\rm trk}$ and $z_{0,\rm trk}$ are the
pseudorapidity and longitudinal impact parameter of the charged track at the position $\vec{x}_{\rm trk}$, respectively.
\begin{figure*}[!htb]
\begin{center}
\includegraphics[width=0.45\textwidth]{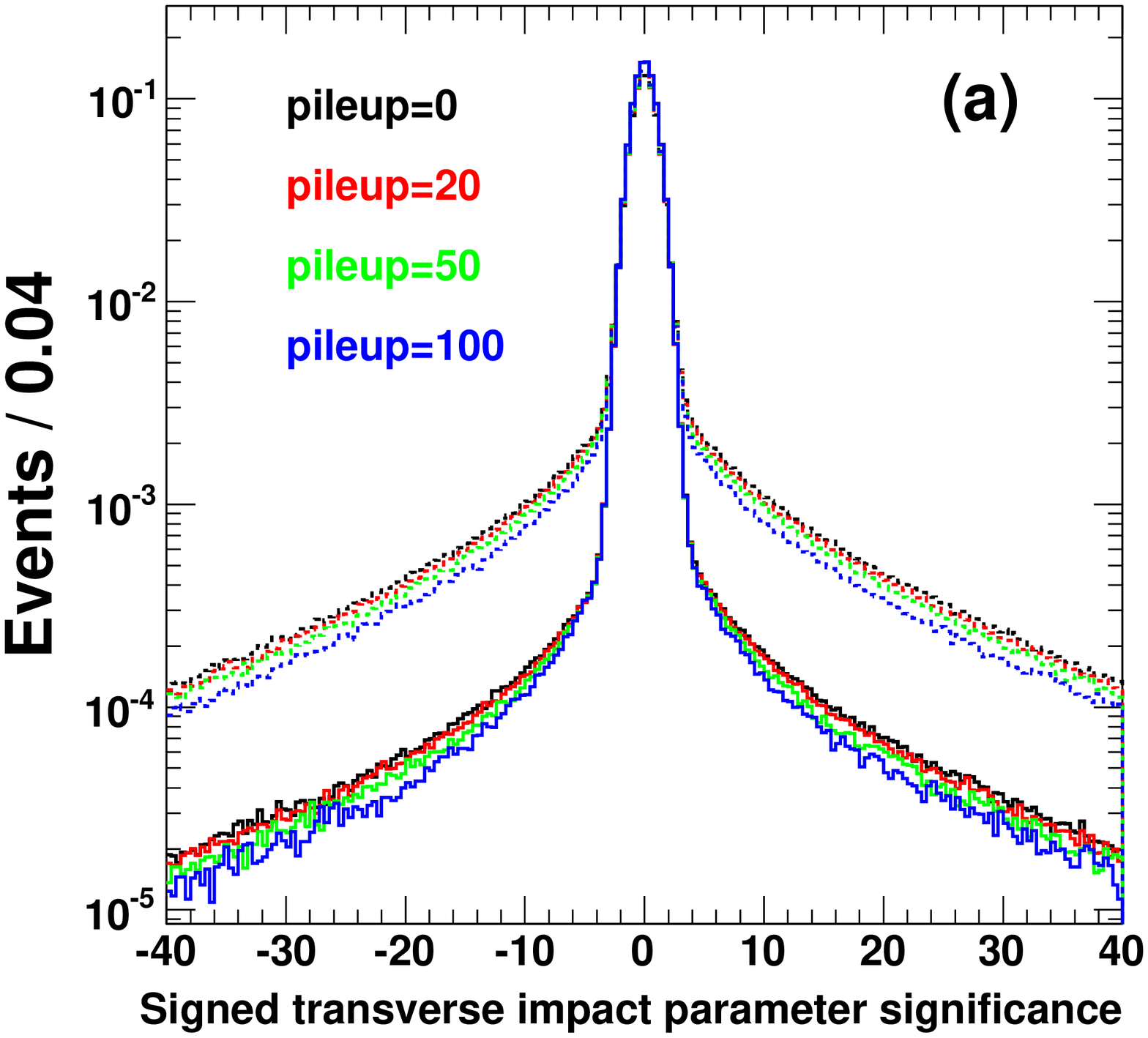}
\includegraphics[width=0.45\textwidth]{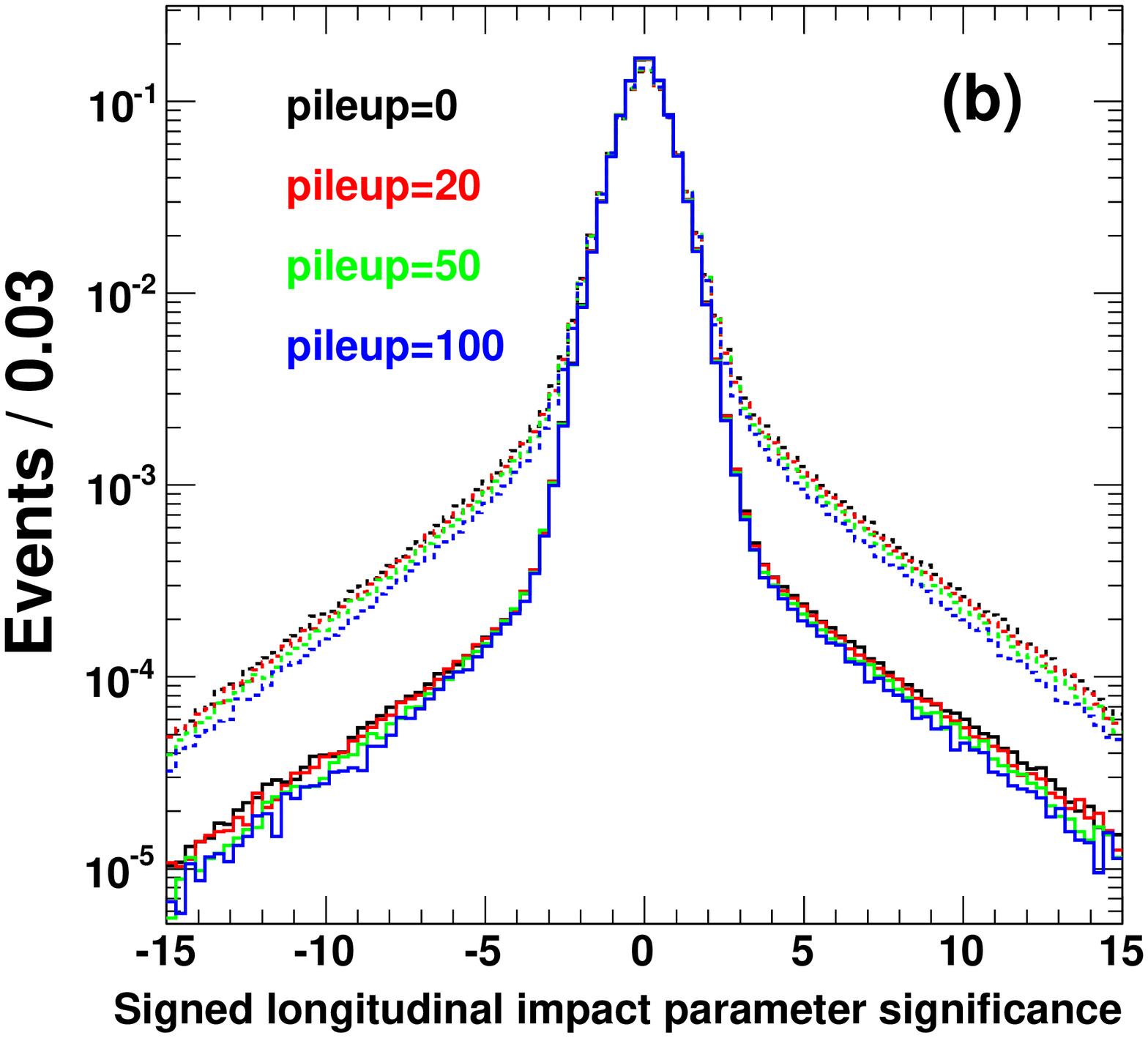}
\includegraphics[width=0.45\textwidth]{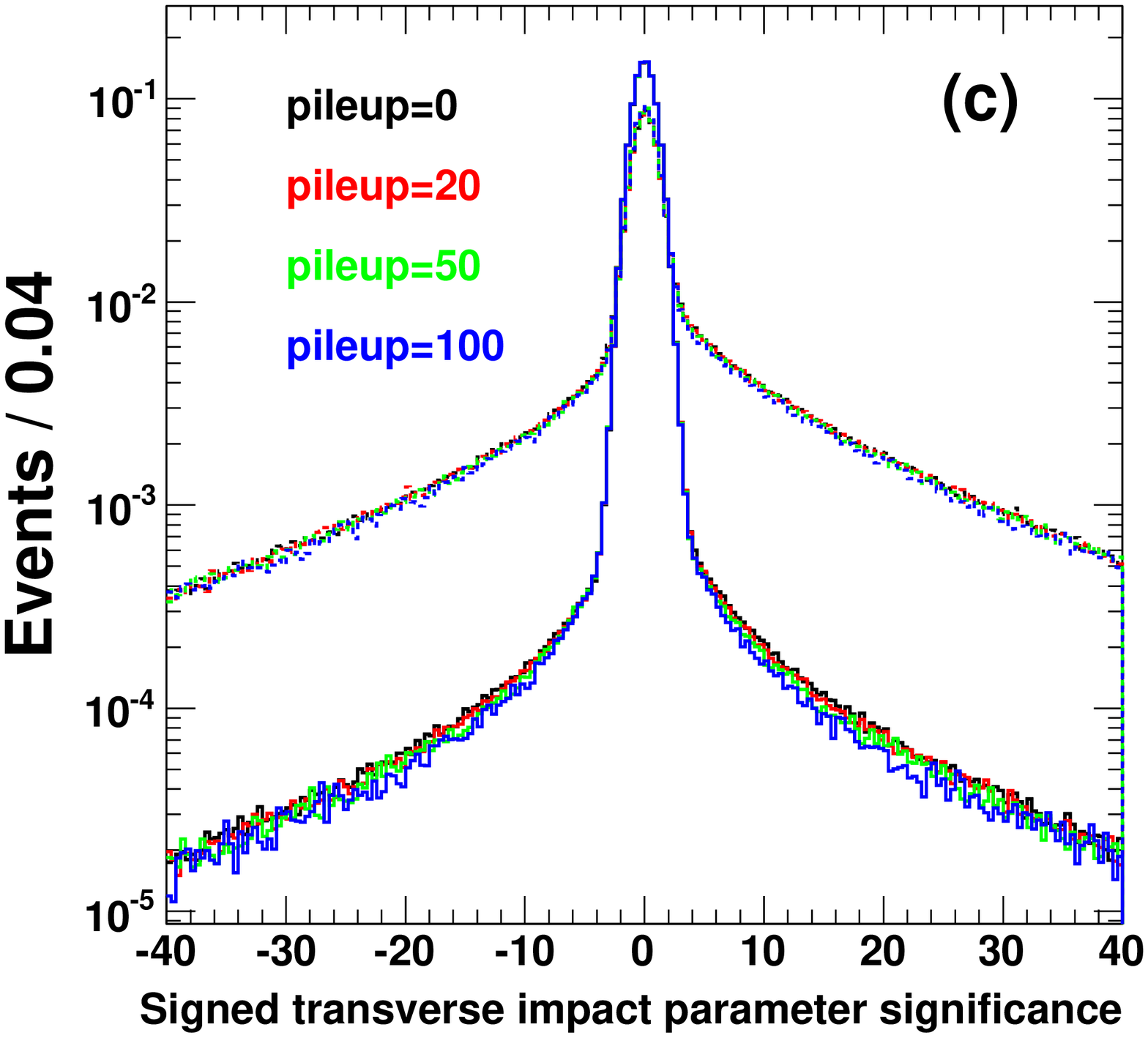}
\includegraphics[width=0.45\textwidth]{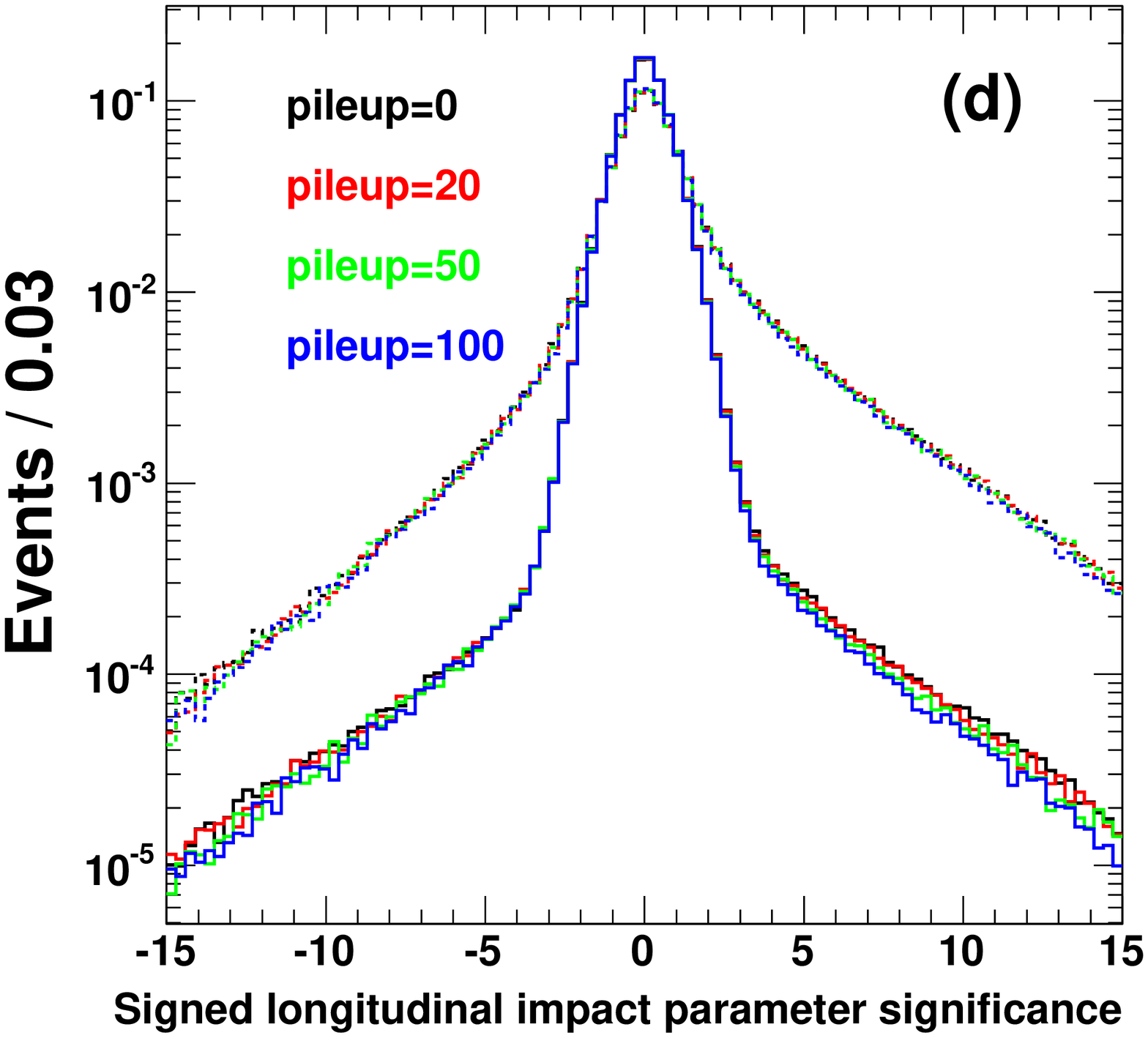}
\caption{The signed transverse impact parameter significance $d_0/\sigma_{{d_0}}$ and 
longitudinal impact parameter significance $z_0/\sigma_{{z_0}}$
under different pileup conditions, where $\sigma_{d_0}$ and $\sigma_{z_0}$ 
are defined as the experimental uncertainties of the measured impact parameter
$d_0$ and $z_0$, respectively.
In (a) and (b), the solid (dashed)  
lines represent the distributions of the charged tracks from the QCD (signal $t$) jets.
In (c) and (d),  the solid (dashed) lines represent the distributions of the charged tracks associated with the subjets ($b$ subjets) in the jet rest frame 
from QCD (signal $t$) jets. All the distributions are normalized to unity.}
\label{fig:signIP}
\end{center}
\end{figure*}

The distributions of the signed impact parameter significances for tracks in QCD jets and signal $t$ jets 
are shown in Figs~\ref{fig:signIP}~(a) and~(b). The significance is defined as the ratio between the impact parameter
and its uncertainty $\sigma$. While we can clearly see a much higher fraction of tracks from the signal $t$ jets with larger
impact parameter significance than the ones from the QCD jets,
the distributions are rather symmetrical and it is contradictory to the expectation and observation in the 
typical $b$ jet tagging algorithm, where the impact parameter distributions from tracks associated with $b$ jets tend to have
positive signs~\cite{Aad:2009wy}, while the experimental resolution generates a random sign for the tracks originating from the primary vertex. 
Studies show that the loss of the sign correlation  is caused by the
mismeasurement of the $b$ quark direction. Unlike a typical $b$ jet, the direction of the $t$ jet is different from 
the $b$ quark direction inside. This correlation is further reduced by the inclusion of the charged tracks generated
by the $W$ bosons in the $t$ jets. 

The identification can be significantly enhanced using the jet rest frame algorithm. We boost all the tracks associated with 
a $t$ jet candidate back to the center-of-mass frame of the jet.  A charged track is considered to be associated with a subjet only if their
angular separation is less than 0.6 in the jet rest frame. By doing so, we separate the tracks that originate from different partons
of the top quark and reject many tracks from underlying events and pileup. The  impact parameters of the tracks associated with
a subjet are then calculated using the subjet momentum in the lab frame. The distributions of the signed impact parameter significances 
for the tracks associated with subjets in QCD jets and subjets containing a $b$ quark ($b$ subjets) in signal $t$ jets are shown in Fig~\ref{fig:signIP}~(c) and~(d).
Their differences are much more predominant comparing to the ones before using the jet rest frame method. 
The impact parameter significances of the tracks associated with a
$b$ subjet in the $t$ jet rest frame show a much larger fraction of positive tail distributions. 

\begin{figure*}[!htb]
\begin{center}
\includegraphics[width=0.45\textwidth]{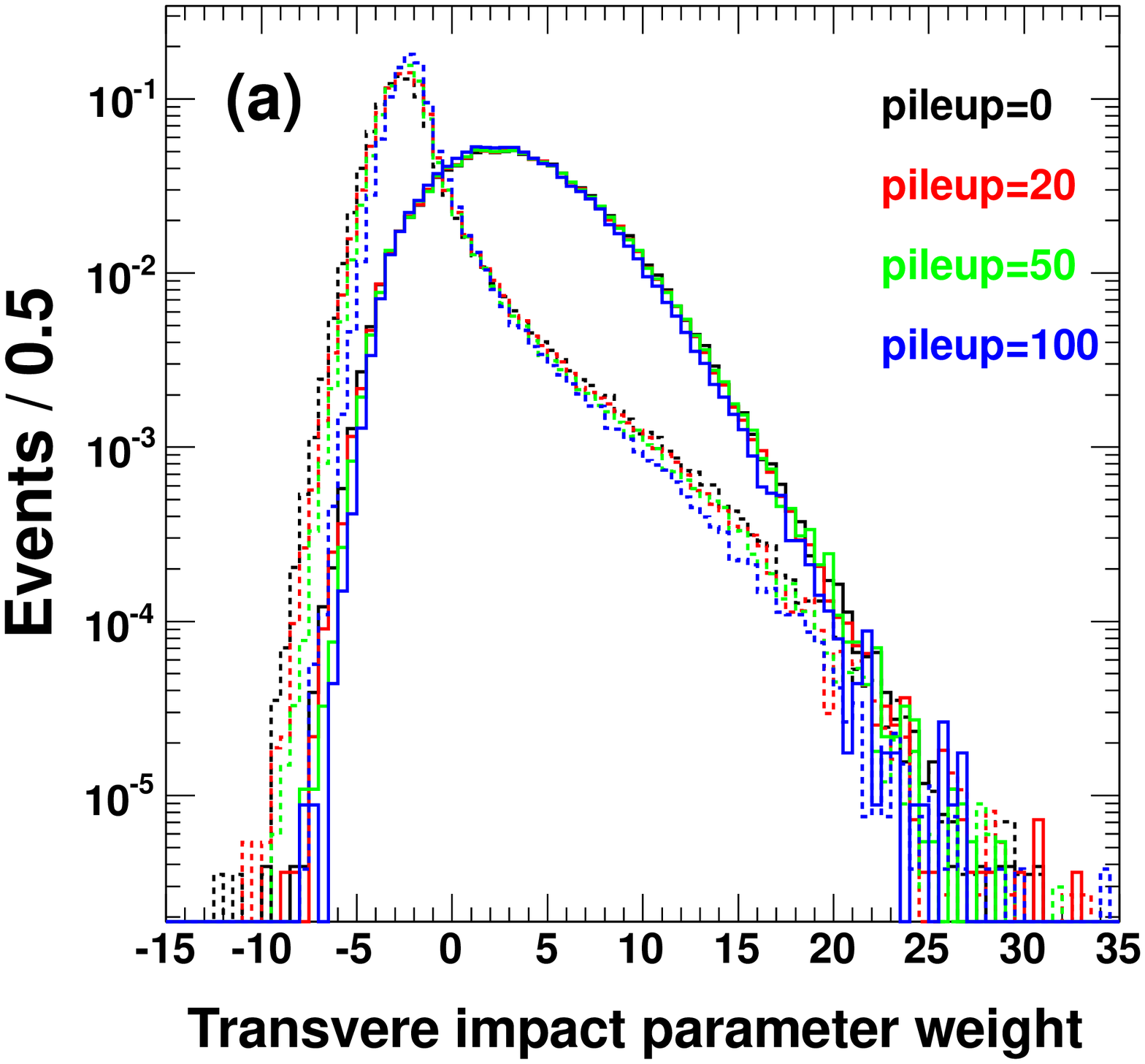}
\includegraphics[width=0.45\textwidth]{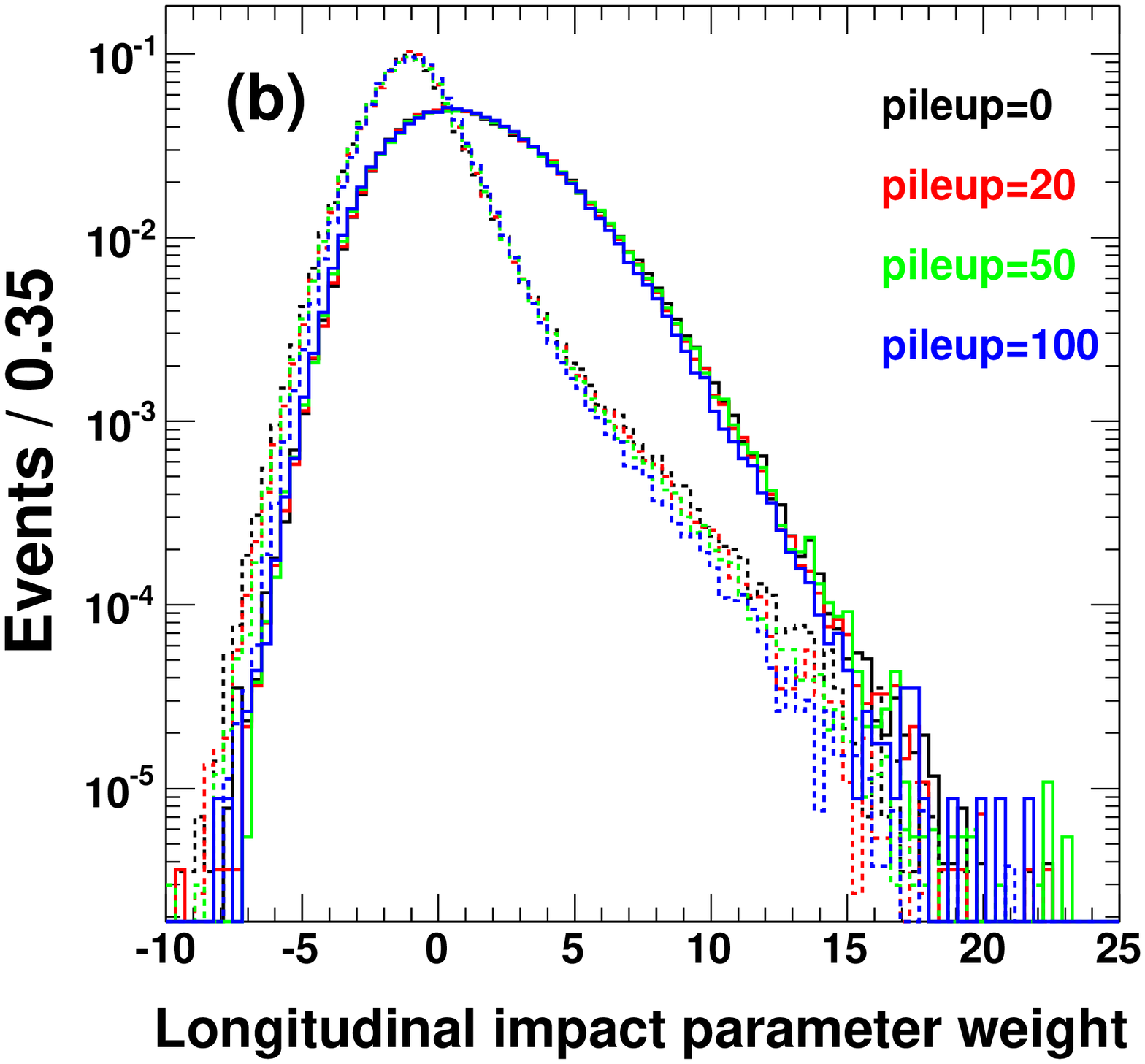}
\includegraphics[width=0.45\textwidth]{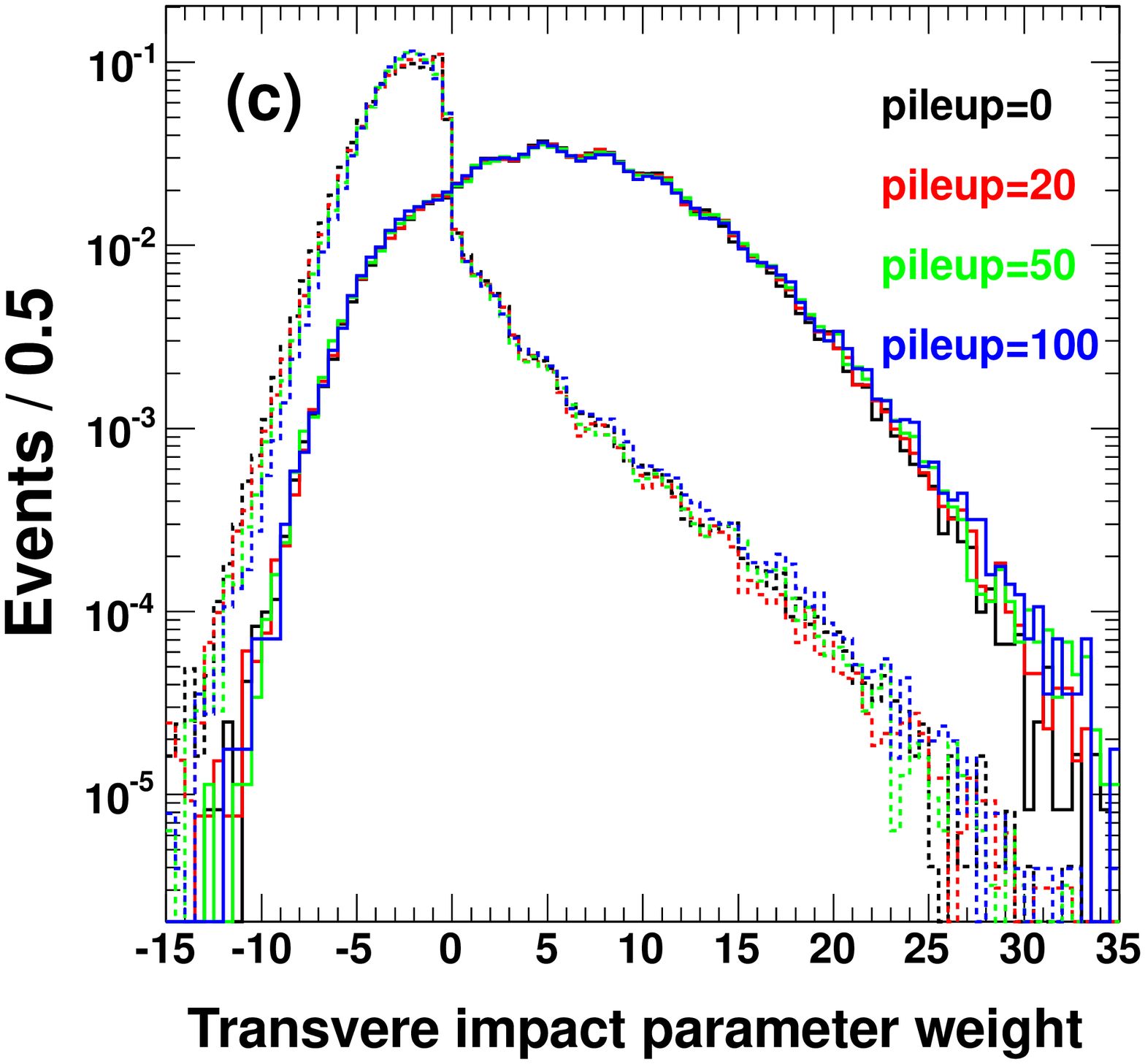}
\includegraphics[width=0.45\textwidth]{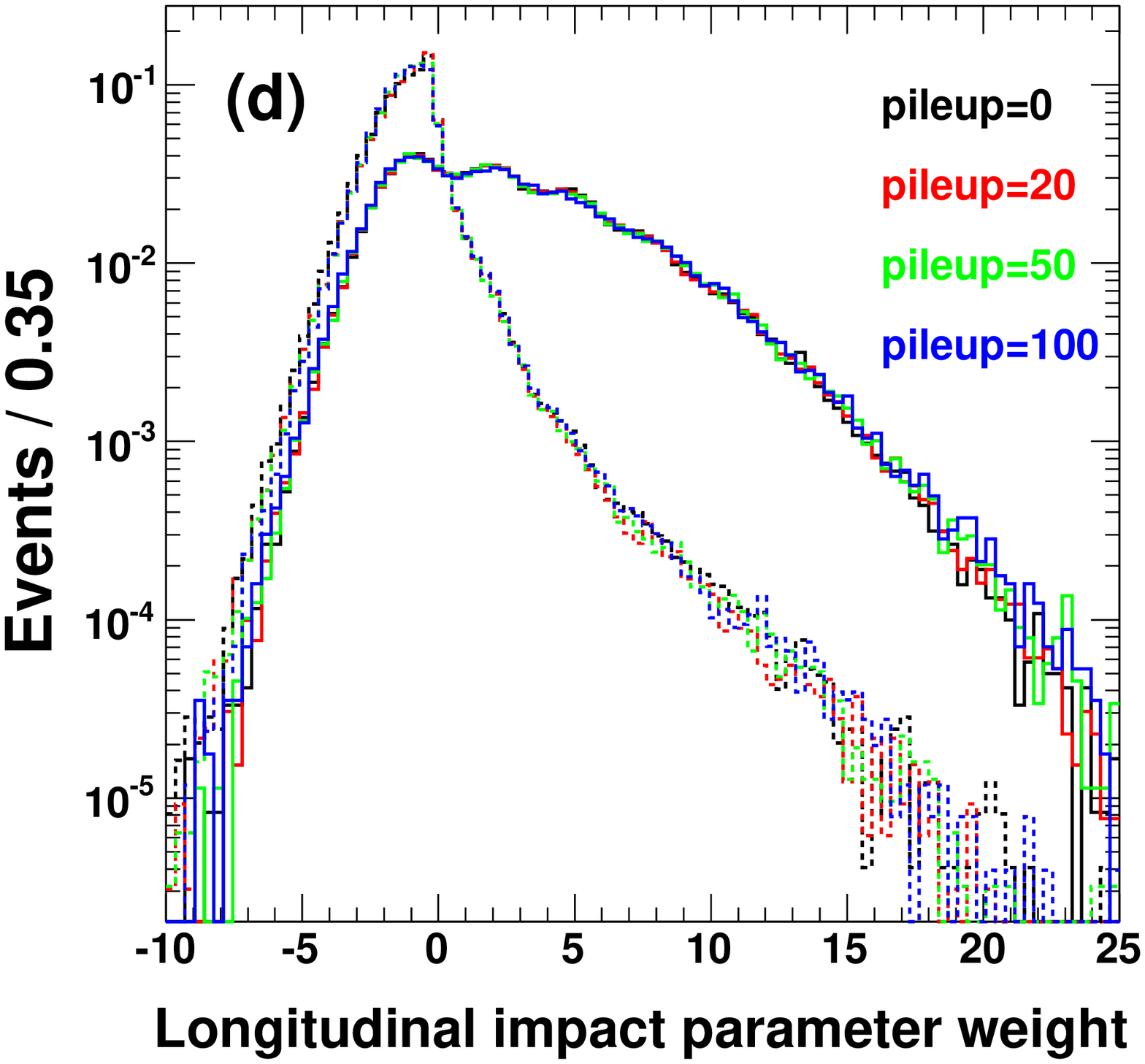}
\caption{The jet weight distributions of transverse and longitudinal impact parameter significances under
different pileup conditions. 
In (a) and (b), the solid (dashed) lines represent the distributions of the charged track associated with
QCD (signal $t$) jets. In (c) and (d), the solid (dashed) lines represent the distributions of the charged tracks associated
with non-$b$ ($b$) subjets in the jet reset frame. All the distributions are normalized to
unity.}
\label{fig:IPweight}
\end{center}
\end{figure*}

We form a likelihood of the tracks associated with a jet. The measured impact parameter significances $S_i$
of the $i$th track in a jet are compared to predefined functions for both $b$ jet and non-$b$ jet hypothesis, $b(S_i)$
and $u(S_i)$, where $b(S)$ and $u(S)$ are the smoothed and normalized distributions of the charged tracks
that are associated with $b$ subjets in the signal $t$ jets and the subjets in the QCD jets, respectively.
The ratio of the probabilities $b(S_i)/u(S_i)$ defines a weight $W_i$. A jet weight $W_{\rm jet}$ is then computed 
as the sum of the $W_i$ from all the tracks associated with the subjet. In case there are no tracks associated with a subject,
its jet weight is assigned to be zero.  For comparison, we also calculate the jet weights for
QCD jets and signal $t$ jets using all the associated tracks without applying subjet reclustering in the jet rest frame.
The distributions of jet weights are shown in Fig~\ref{fig:IPweight}. Again, the signal and background distribution calculated
using subjet information in the jet rest frame show much more significant separations.

\begin{figure*}[!htb]
\begin{center}
\includegraphics[width=0.45\textwidth]{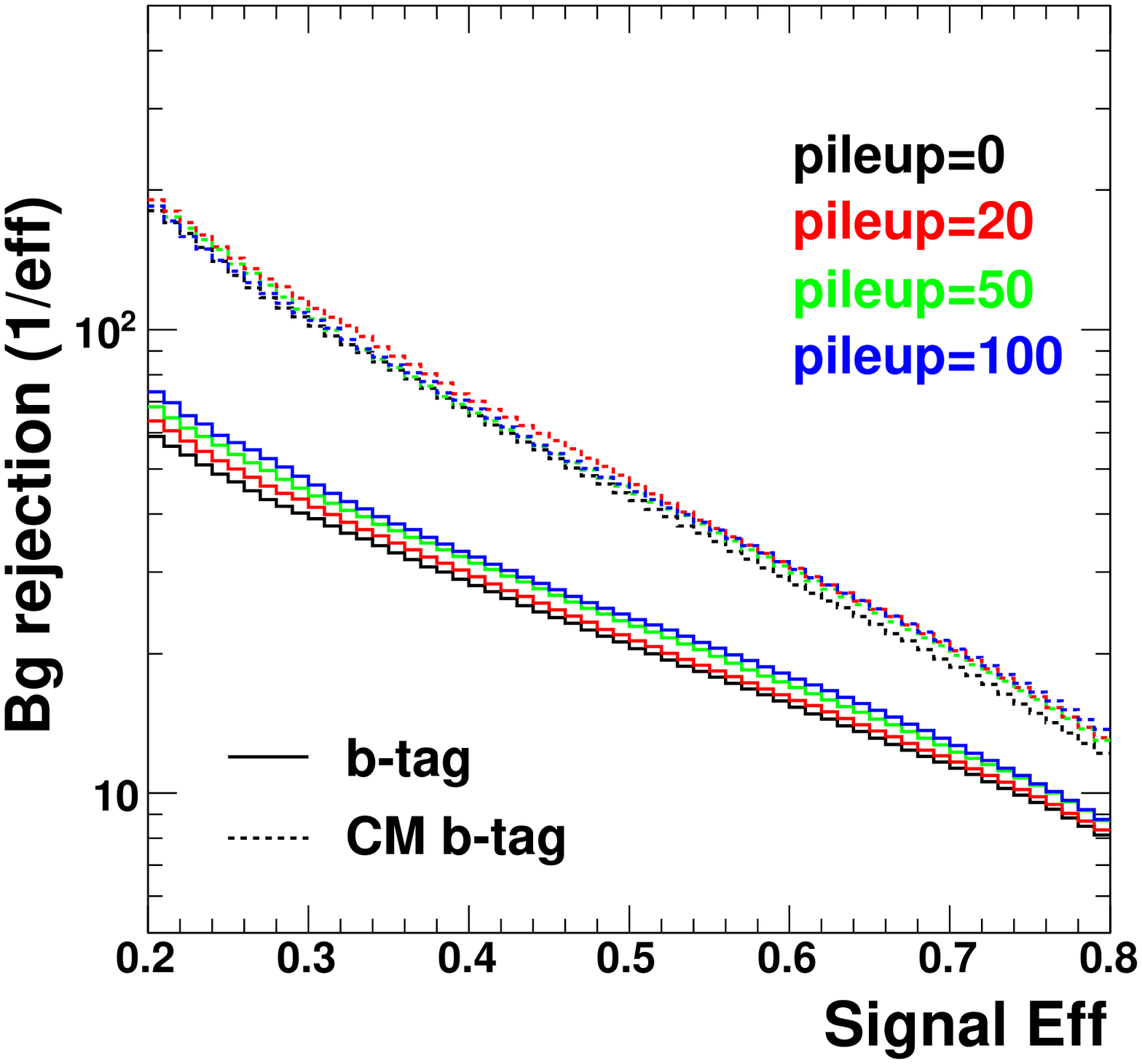}
\includegraphics[width=0.45\textwidth]{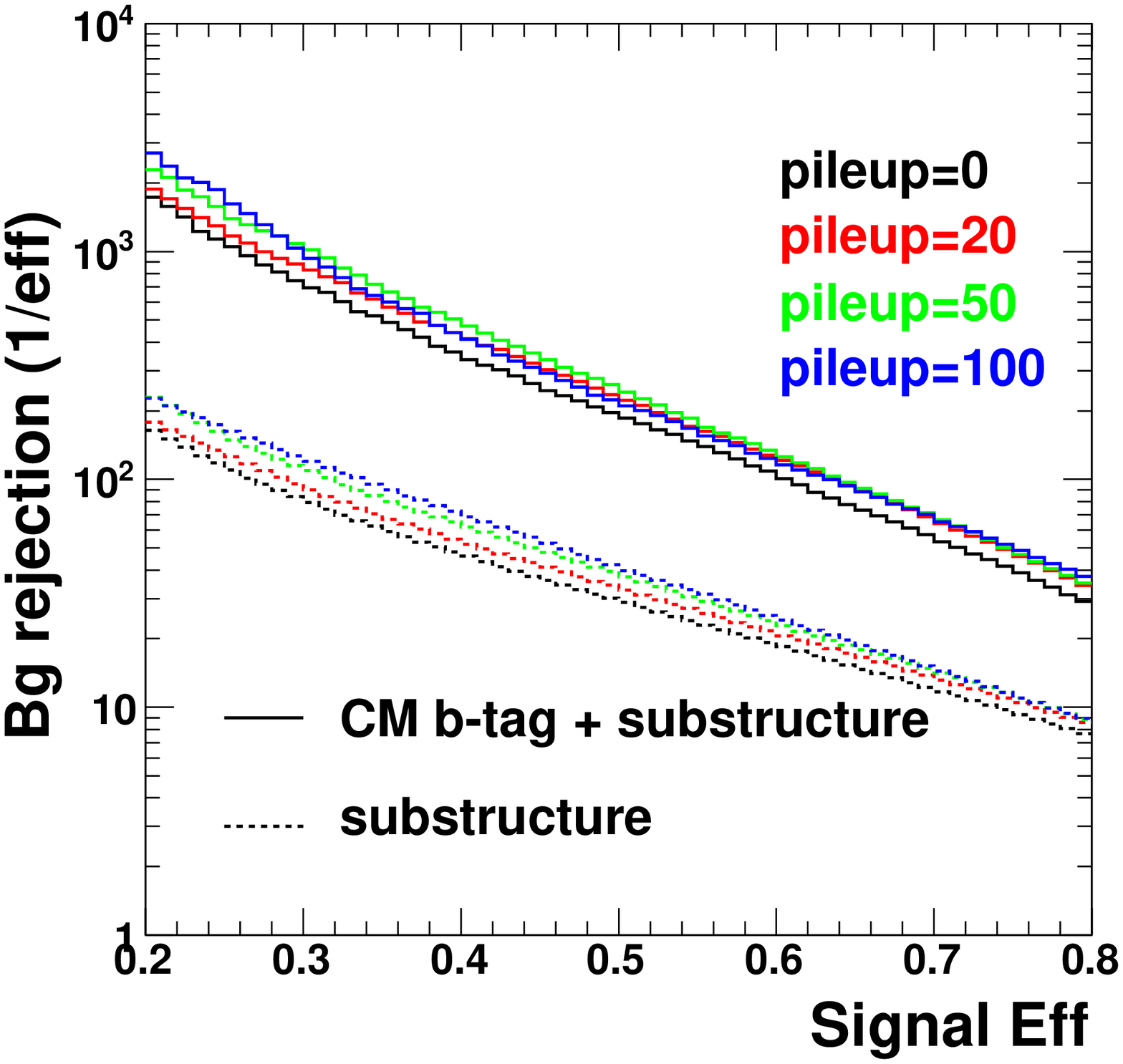}
\caption{The background rejection of QCD jets vs. the signal efficiency of $t$ jets for the top tagger in  
different pileup conditions. Left: the top identification based on a regular $b$-tagging (solid) method and the
method in the jet rest frame (dashed). Right: the top identification using jet substructure with (solid) and without (dashed)
$b$ tagging in the jet rest frame.}
\label{fig:EffvsRej}
\end{center}
\end{figure*}

The final $b$ quark identification variable is constructed using a boosted decision tree (BDT) algorithm 
with the jet weights of  the 3 leading subjets in the jet rest frame  in order to take into account their correlations. 
In order to compare to the application of typical $b$-tagging algorithms on $t$ jets, we also construct a BDT variable using the jet weights that
are calculated with all the associated charged tracks. The signal efficiency of $t$ jets by identifying the $b$ quark inside 
vs. the background rejection of QCD jets for the BDT variable is shown in Fig.~\ref{fig:EffvsRej}. Regardless of the pileup conditions, 
the $b$ quark identification method in the jet rest frame we propose can easily reduce the contribution
of the QCD jet background by approximately 100, with only a factor of three reduction for the $t$ jet signal identification efficiency.
Its performance is a few times better than the direct application of typical $b$-tagging on boosted $t$ jets. 
As shown in Fig.~\ref{fig:EffvsRej}, the performance of the tagger is generally slightly better with higher pileup. 
Studies show that this is an effect that is caused
by the selection of jets used in the evaluation of the $b$-tagging performance. In our studies, we use only jets that
have $p_{\rm T}> 600\,\gev$, $50\,\gev <m_{\rm jet}<350\,\gev$ and at least 3 subjets with $E_{\rm jet} > 10\,\gev$ in its rest frame.
As a result, when pileup increases, many QCD jets that
otherwise would not satisfy the jet selection criteria are
selected.

The addition of the identification of a $b$ quark inside $t$ jets can be used to improve existing 
top taggers~\cite{Altheimer:2012mn} that are solely used on jet substructure information. Here we demonstrate 
such an application by combining the jet weights in the jet rest frame with the other jet substructure variables 
introduced in Ref.~\cite{Chen:2013ola}, such as the energies of the three leading subjets, the invariant mass of each
two subjet combinations, the energy asymmetry between $W$ candidate and $b$ jet candidate in the jet rest farme,
as well as the opening angle between them. A BDT variable is subsequently formed using the variable described above.
As shown in Fig.~\ref{fig:EffvsRej}, the background rejection achieved by the new top tagger based on both 
$b$ identification and jet substructure in the jet rest frame is  almost an order of magnitude 
higher compared to the ones that only rely on the jet substructure information~\cite{Chen:2013ola,Altheimer:2012mn}. 
This observation implies that 2 orders of magnitude further reduction  can be achieved for the dominant QCD
background in the searches for new heavy resonance decaying to a $t\bar{t}$ final state, and therefore greatly 
improve the expected experimental sensitivities.

In conclusion, we study the identification of
the $b$ quark inside boosted hadronically decaying top quark in the center-of-mass frame of the jet. 
We demonstrate that the method can greatly reduce the QCD jet background while maintaining a high
identification efficiency of the boosted top quark even under  a very large pileup condition. We compare the method to the commonly used
$b$-tagging algorithm in a jet, and show that our method has a much smaller fake rate  of the QCD jets 
for the same efficiency.  When combining the $b$ tagging in the center-of-mass 
frame of the top jet with the jet substructure information, we can improve the rejection rate of QCD jet 
background by almost an order of magnitude while maintaining the same identification efficiency of the boosted top quark. 
The study shows a  good prospect for the search for heavy mass particles in the decay channels containing 
$t$ quarks with the LHC experiments at 14\,\tev\ center-of-mass energy. 

We thank  Soeren Prell for many discussions and valuable comments on the manuscript. 
This work is supported by
the Office of Science of the U.S. Department of Energy
under Contracts No. DE-FG02-12ER41827 and No. DE-FG02-13ER42027.


\begin{thebibliography}{99}
\bibitem{Agashe:2006hk} 
  K.~Agashe, A.~Belyaev, T.~Krupovnickas, G.~Perez and J.~Virzi;
  Phys.\ Rev.\ D {\bf 77}, 015003 (2008).
  R.~Contino, T.~Kramer, M.~Son and R.~Sundrum,
  JHEP {\bf 0705}, 074 (2007);
  S.~Matsumoto, M.~M.~Nojiri and D.~Nomura,
  Phys.\ Rev.\ D {\bf 75}, 055006 (2007);
  A.~L.~Fitzpatrick, J.~Kaplan, L.~Randall and L.~T.~Wang,
  JHEP {\bf 0709}, 013 (2007);
  G.~Altarelli, B.~Mele and M.~Ruiz-Altaba,
  Z.\ Phys.\ C {\bf 45}, 109 (1989)
  [Erratum-ibid.\ C {\bf 47}, 676 (1990)].      

\bibitem{Aad:2012wm} 
[ATLAS Collaboration],
  Eur.\ Phys.\ J.\ C {\bf 72}, 2083 (2012).

\bibitem{Aad:2012dpa} 
[ATLAS Collaboration],
  JHEP {\bf 1209}, 041 (2012).

\bibitem{Aad:2012raa} 
[ATLAS Collaboration],
  JHEP {\bf 1301}, 116 (2013).

\bibitem{Aad:2012ej} 
[ATLAS Collaboration],
Phys.\ Rev.\ Lett.\  {\bf 109}, 081801 (2012).
  
\bibitem{Chatrchyan:2012ku} 
[CMS Collaboration],
  JHEP {\bf 1209}, 029 (2012).
      
\bibitem{Chatrchyan:2012cx} 
[CMS Collaboration],
  JHEP {\bf 1212}, 015 (2012).
   
\bibitem{Chatrchyan:2012yca} 
[CMS Collaboration],
  Phys.\ Rev.\ D {\bf 87}, 072002 (2013).
     
\bibitem{Chatrchyan:2012gqa} 
[CMS Collaboration],
  Phys.\ Lett.\ B {\bf 718}, 1229 (2013).
          
\bibitem{Thaler:2008ju}
  J.~Thaler and L.~T.~Wang,
  J. High Energy Phys. {\bf 0807}, 092 (2008).

\bibitem{Kaplan:2008ie}
  D.~E.~Kaplan, K.~Rehermann, M.~D.~Schwartz and B.~Tweedie,
  Phys.\ Rev.\ Lett.\  {\bf 101}, 142001 (2008).

\bibitem{Almeida:2008tp}
  L.~G.~Almeida, S.~J.~Lee, G.~Perez, I.~Sung and J.~Virzi,
  Phys.\ Rev.\  D {\bf 79}, 074012 (2009).

\bibitem{Krohn:2009wm}
  D.~Krohn, J.~Shelton and L.~T.~Wang,
  J. High Energy Phys. {\bf 1007}, 041 (2010).

\bibitem{Chekanov:2010vc}
  S.~Chekanov and J.~Proudfoot,
  Phys.\ Rev.\  D {\bf 81}, 114038 (2010).

\bibitem{Plehn:2010st}
  T.~Plehn, M.~Spannowsky, M.~Takeuchi, D.~Zerwas,
  J. High Energy Phys. {\bf 1010}, 078 (2010).

\bibitem{Bhattacherjee:2010za}
  B.~Bhattacherjee, M.~Guchait, S.~Raychaudhuri and K.~Sridhar,
  Phys.\ Rev.\  D {\bf 82}, 055006 (2010).

\bibitem{Rehermann:2010vq}
  K.~Rehermann, B.~Tweedie,
  J. High Energy Phys. {\bf 1103}, 059 (2011).

\bibitem{Chekanov:2010gv}
  S.~Chekanov, C.~Levy, J.~Proudfoot and R.~Yoshida,
  Phys.\ Rev.\  D {\bf 82}, 094029 (2010)

\bibitem{Jankowiak:2011qa}
M.~Jankowiak and A.~J.~Larkoski,
JHEP {\bf 1106}, 057 (2011).

\bibitem{Thaler:2011gf}
 J.~Thaler and K.~Van Tilburg,
JHEP {\bf 1202}, 093 (2012).

\bibitem{Soper:2012pb}
  D.~E.~Soper and M.~Spannowsky,
    Phys.\ Rev.\ D {\bf 87}, 054012 (2013).
  
\bibitem{Chen:2013ola} 
  C.~Chen,
  Phys.\ Rev.\ D {\bf 87}, 074007 (2013).
  
    
\bibitem{Chen:2011ah} 
  C.~Chen,
  Phys.\ Rev.\ D {\bf 85}, 034007 (2012).

\bibitem{Sjostrand:2006za}
  T.~Sj{\"o}strand, S.~Mrenna, P.~Z.~Skands,
 J. High Energy Phys. {\bf 0605}, 026 (2006).

\bibitem{Aad:2008zzm} 
[ATLAS Collaboration],
  JINST {\bf 3}, S08003 (2008).
  

\bibitem{Cacciari:2005hq} 
  M.~Cacciari and G.~P.~Salam,
  Phys.\ Lett.\ B {\bf 641}, 57 (2006).

    
\bibitem{Cacciari:2008gp}
M.~Cacciari, G.~P.~Salam and G.~Soyez,
  J. High Energy Phys. {\bf 0804}, 063 (2008).

 
 \bibitem{Dokshitzer:1997in} 
  Y.~L.~Dokshitzer, G.~D.~Leder, S.~Moretti and B.~R.~Webber,
  JHEP {\bf 9708}, 001 (1997).
  
\bibitem{Aad:2009wy} 
[ATLAS Collaboration],
  arXiv:0901.0512 [hep-ex].
    
 \bibitem{Altheimer:2012mn} 
  A.~Altheimer {\it et al.},
  J.\ Phys.\ G {\bf 39}, 063001 (2012).
   
\end{thebibliography}
\end{document}